\documentclass[rmp,twocolumn,floatfix,authordate1-4,amssymb]{revtex4}

\usepackage{times}
\input psfig.sty
\usepackage{graphicx}  
                       
\begin{document}

\title{Lessons of Slicing Membranes: Interplay of Packing, Free Area, and 
       Lateral Diffusion in Phospholipid\,/\,Cholesterol Bilayers} 

\setcounter{page}{1} 
 
\author{Emma Falck}
\affiliation{Laboratory of Physics and Helsinki Institute of Physics,
Helsinki University of Technology, P.\,O. Box 1100, FIN--02015 HUT, Finland}

\author{Michael Patra}
\author{Mikko Karttunen}
\affiliation{Biophysics and Statistical Mechanics Group,
Laboratory of Computational Engineering, Helsinki University
of Technology, P.\,O. Box 9203, FIN--02015 HUT, Finland}

\author{Marja T. Hyv\"onen} 
\affiliation{Wihuri Research Institute, Kalliolinnantie 4, 
FIN--00140 Helsinki, Finland, and 
Laboratory of Physics and Helsinki Institute of Physics, 
Helsinki University of Technology, P.\,O. Box 1100, FIN--02015 HUT, Finland} 

\author{Ilpo Vattulainen}
\affiliation{Laboratory of Physics~and~Helsinki Institute of Physics,
Helsinki University of Technology, P.\,O. Box 1100, FIN--02015 HUT, Finland}

\date{February 09, 2003}

\begin{abstract}
We employ 100\,ns molecular dynamics simulations to study the 
influence of cholesterol on structural and dynamic properties 
of dipalmitoylphosphatidylcholine (DPPC) bilayers in the fluid phase. 
The effects of the cholesterol content on the bilayer 
structure are considered by varying the cholesterol 
concentration between 0 and 50\,\%. We concentrate on the free 
area in the membrane and investigate quantities that are 
likely to be affected by changes in the free area and free volume 
properties. It is found that cholesterol has a strong impact 
on the free area properties of the bilayer. The changes in the 
amount of free area are shown to be intimately related to alterations 
in molecular packing, ordering of phospholipid tails, and compressibility. 
Further, the behavior of the lateral diffusion of both DPPC
and cholesterol molecules with an increasing amount of cholesterol 
can in part be understood in terms of free area. Summarizing,
our results highlight the central role of free area in comprehending
the structural and dynamic properties of membranes 
containing cholesterol. 
\end{abstract}

\maketitle

\section{Introduction}

Cholesterol is one of the most prominent molecular species 
in the plasma membranes of mammalian cells. It is a tremendously 
important molecule, a component essential for the very existence and 
multiplication of cells~\cite[and references therein]{Fin93,Ohv02}. 
It is abundant in the plasma membranes of higher organisms: 
depending on the exact lipid composition, the plasma membrane 
may contain of the order of 20\,--\,50\,\% cholesterol~\cite[]{Alb94}.

Eukaryotic cells do not seem to be able to grow and differentiate 
properly without cholesterol. It has been firmly established that 
cholesterol modulates the physical properties of the plasma 
membrane~\cite[]{McM96}. A finite cholesterol content has been said 
to improve the characteristics of a simple phospholipid bilayer 
and allow for wider variations in the lipid composition of the 
membrane~\cite[]{Vis90}. Perhaps not surprisingly, cholesterol 
is one of the primary molecules in lipid 
rafts~\cite[and references therein]{Edi03,Sil03,Sim97}, 
i.\,e., microdomains rich in cholesterol, sphingomyelin, and 
saturated phospholipids. Rafts have been thought to
confine proteins involved in e.g.~signal transduction events, 
and hence act as platforms for adhesion and signaling.
Consequently, one could well imagine that as cholesterol alters 
the properties of the bilayer, it might affect the functioning 
of the embedded proteins~\cite[]{Can99,Yea91}.

The effects of cholesterol on the properties of phospholipid 
bilayers are diverse. In the physiologically relevant fluid 
phase, adding cholesterol to the bilayer leads to increased 
orientational order in the phospholipid 
tails~\cite[]{Chi02,Hof03,McM96,San90b} and smaller average 
areas per molecule~\cite[]{Pet99}. In other words, cholesterol 
modifies the packing of molecules in bilayers. Other important 
effects are changes in passive permeability of small solutes 
\cite[and references therein]{Jed03b,Xia99} and suppressed 
lateral diffusion of phospholipids in bilayers with cholesterol 
\cite[]{Alm92,Gal79,Hof03,Pol01,Vat03}. Both permeability and 
lateral diffusion, in turn, are strongly affected by the amount 
and distribution of free volume or area in a membrane, 
i.\,e.,~space not occupied by phospholipids, cholesterols, 
or water. Cholesterol thus seems to simultaneously 
influence packing, free area, diffusion, and permeability in lipid 
bilayers, and it is reasonable to expect that the changes in these 
properties are somehow coupled.

Although there is a wealth of information on the effects of 
cholesterol on lipid bilayers, the interplay of packing, free 
area, diffusion, and permeability has not yet been studied 
systematically. Experimental electron density profiles~\cite[]{McI78} 
and deuterium nuclear magnetic resonance (NMR) data~\cite[]{San90b} 
suggest that cholesterol should influence the packing inside 
membranes. Fluorescence recovery after photobleaching (FRAP) 
experiments, in turn, have been used to study the dependence of 
lateral diffusion coefficients on free area~\cite[]{Alm92}. More 
information at the atomic level, however, is essential for gaining 
a detailed understanding of the effect of cholesterol on lipid 
bilayers. Such atomic-level information can be obtained from computer 
simulations. Molecular dynamics in particular provides a unique 
tool to investigate both the structure and dynamics of lipid 
membranes with a level of detail missing in any experimental 
technique. Until recently, however, systematic simulation studies 
have been limited by the extensive computational requirements.

In the present study, we investigate the cholesterol-induced changes 
in packing, free area, ordering, and lateral diffusion in phospholipid 
bilayers. Specifically, we study the presumptive interplay between 
these changes. To this end, we employ 100\,ns molecular dynamics 
simulations on dipalmitoyl phosphatidylcholine (DPPC)\,/\,cholesterol 
bilayers, with cholesterol concentrations ranging from 0 to 50 mol\,\%. 
While detailed multi-nanosecond simulation studies on atomic level 
have emerged only very recently~\cite[]{Hof03,Sco02,Tie97}, there 
exist large amounts of experimental studies for DPPC\,/\,cholesterol 
bilayers~\cite[and references therein]{McM96,San90a,San90b,Vis90}. 
These previous studies and the experimental results in particular 
offer us an excellent platform for comparison.

In order to further enhance the understanding of the effect of 
cholesterol on bilayers, we introduce a novel method for investigating
the packing and free area in bilayers. The scope of this technique is 
very wide. It allows us to estimate how much space DPPC, cholesterol, 
and water molecules on average occupy in different regions of 
the bilayer. Consequently, it yields information on the amount 
and location of {\it free space} in the bilayer. As discussed
below, this is related to various structural aspects such as the 
ordering of lipids in a membane. Our method also 
provides valuable insight into dynamic properties. For example, 
our approach allows us to determine the area compressibility 
across a membrane, and hence yields information on rate-limiting regions 
for lateral diffusion. In addition, as the method enables us  
to examine changes in free area with an increasing cholesterol 
content, we may estimate diffusion coefficients
in terms of free area theories for lateral diffusion. The 
present approach can be applied to a wide range of 
different kinds of membrane systems, including one- and 
multi-component bilayers, and bilayers with embedded solutes, 
probes, and proteins.

We find that cholesterol strongly affects the amount of 
space occupied by molecules in different parts of a 
phospholipid bilayer. The close-packed areas 
occupied by the tails of DPPC molecules 
can be explained by the ordering of the tails, and a 
simple relation~\cite{Pet99} can be used for quantifying 
the dependence of close-packed area on ordering. 
The amount and location of free space is significantly reduced 
by an increasing cholesterol content,
and clearly reflect the total space occupied by DPPC and 
cholesterol molecules. The lateral diffusion coefficients, too,
show a substantial decrease with an increasing cholesterol 
concentration. We find that so-called free area 
theories~\cite[]{Alm92,Coh59,Gal79}, 
which are essentially two-dimensional mean-field models, 
correctly predict this reduction, but are not applicable
to quantitatively describing lateral diffusion in lipid bilayers.

\section{Model and Simulation Details}

{\small
We studied fully hydrated lipid bilayer systems consisting 
of 128 molecules, i.\,e.,~DPPC's and cholesterols, and 3655 
water molecules. Since the main focus of this paper is on studying the
effects of cholesterol on phospholipid bilayers, we were interested in
bilayers with varying amounts of cholesterol. To this end, we studied 
a pure DPPC bilayer and composite DPPC\,/\,cholesterol bilayers 
with six different cholesterol molar fractions: $\chi = 0$\,\%, 
4.7\,\%, 12.5\,\%, 20.3\,\%, 29.7\,\%, and 50.0\,\%.

The starting point was a united atom model 
for a fully hydrated pure DPPC bilayer that has been validated 
previously~\cite[]{Tie96,Pat03a}. The parameters for bonded and 
non-bonded interactions for DPPC molecules were taken from 
a study of a pure DPPC bilayer~\cite[]{Ber97} available at 
{\tt http://moose.\-bio.\-ucalgary.\-ca/\-Downloads/\-lipid.itp}. 
The partial charges are from the underlying model 
description~\cite[]{Tie96} and can be found at 
{\tt http://moose.\-bio.\-ucalgary.\-ca/\-Downloads/\-dppc.itp}. 
For water, the SPC model~\cite[]{Ber81} was used. As our initial 
configuration for the pure DPPC bilayer we used the final 
structure of run E discussed in Ref.~\cite[]{Tie96} and available 
at {\tt http://moose.\-bio.\-ucalgary.\-ca/\-Downloads/\-dppc128.pdb}. 
The bilayer is aligned such that it lies in the xy-plane, i.\,e., the
bilayer normal is parallel to the z-axis.

The cholesterol force field and the initial shape of an individual 
cholesterol molecule were taken from 
{\tt http://www.\-gromacs.\-org/\-topologies/\-uploaded\_molecules/\-
cholesterol.tgz}~\cite[]{Hol01}. Cholesterols were 
introduced to the bilayer by choosing DPPC 
molecules from the pure phospholipid bilayer at random and 
replacing them by cholesterols. The same number of DPPC molecules 
was replaced in each of the two monolayers. In practice, the 
center of mass (CM) of a cholesterol molecule was moved to the 
CM position of the removed DPPC molecule. The main axis
of inertia of each inserted cholesterol was parallel to 
the z-axis, and each molecule was rotated by a random 
angle around the z-axis.

The molecular dynamics (MD) simulations were performed at a temperature
$T = 323$\,K using the GROMACS~\cite[]{Lin01} molecular simulation 
package. The time step for the simulations was chosen to be 2.0\,fs. 
The lengths of all bonds were kept constant with the LINCS 
algorithm~\cite[]{Hes97}. Lennard-Jones interactions were cut 
off at 1.0\,nm without shift or switch functions. Long-range 
electrostatic interactions were handled using the Particle-Mesh 
Ewald~\cite[]{Ess95a} method, which has been shown to be a reliable 
method to account for long-range interactions in lipid bilayer 
systems~\cite[]{Pat03a}. Electrostatic interactions within 1.0\,nm 
were calculated at each time step, while interactions beyond 
this range were determined every ten time steps. These choices 
follow the parametrization of DPPC~\cite[]{Tie96} and correspond 
to a scheme called twin-range cutoff.

After an inital energy minimization, we needed to
equilibrate the system to fill the small voids left by replacing
DPPC molecules by somewhat smaller cholesterol molecules.
The equilibration was commenced by 50\,ps 
of $NVT$ molecular dynamics with a Langevin thermostat
using a coupling time of 0.1\,ps, i.\,e., every 0.1\,ps the
velocities of all particles were completely randomized 
from a Maxwell distribution corresponding to the target temperature.
This complete loss of memory after 0.1\,ps reduces 
the amount of ballistic motion of atoms inside a void.
The equilibration was continued by 500\,ps of $NpT$ molecular
dynamics at a pressure of 1\,bar
with a Langevin thermostat and a Berendsen barostat~\cite[]{Ber84}.
The time constant for the latter was set to 1\,ps,
and the height of the simulation box was allowed to vary separately 
from the cross-sectional area of the box.

Finally, for every cholesterol concentration, we performed 100\,ns 
of MD in the $NpT$ ensemble with a Berendsen thermostat and 
barostat~\cite{Ber84}. The barostat was the same as the one 
described above, and the thermostat was set to separately couple 
the DPPC, cholesterol, and water molecules to a heat bath with 
a coupling time of $0.1~\mathrm{ps}$. The six simulations took
a total of approximately 60,000 hours of CPU time. For all 
systems up to and including the cholesterol molar fraction of 
29.7\,\%, a simulation time of 100\,ns guarantees a good sampling 
of the phase space. The results for $50\,\%$ cholesterol should 
be regarded with some caution, as the diffusion of the DPPC and 
cholesterol molecules is already quite slow, see Section~\ref{sec_diff}. 
As mixing of DPPC and cholesterol molecules in this case is 
quite limited, the system probably bears traces of its initial 
configuration. This applies to all state-of-the-art simulation 
studies of phospholipid\,/\,cholesterol systems, and has been 
mentioned by other authors~\cite[]{Smo99}.}

\section{Results and Discussion}

\subsection{Equilibration \label{sec_eq}}

One of the most important quantities describing lipid bilayers 
is the average area per molecule. The average area per 
molecule for a given configuration, $A$, is computed by
dividing the size of the simulation box in the xy-plane, designated 
$A_\mathrm{tot}$, by  $N$, the total number of molecules, 
i.\,e.,~DPPC's and cholesterols, in a monolayer. The average area per 
molecule can, among other things, be used for monitoring the equilibration
of the membrane.

\begin{figure}[h]
\centering
\includegraphics[width=\columnwidth,bb=105 295 425 535,scale=1,clip=true]{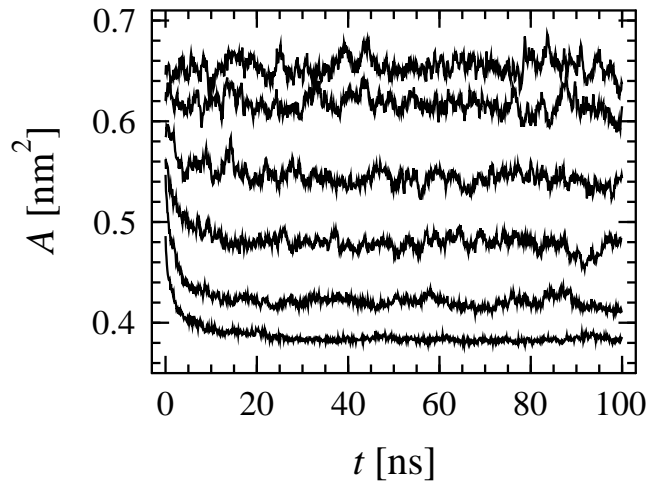}
\caption{Temporal behavior of area per molecule. The curves correspond to, 
from top to bottom, cholesterol concentrations 0.0\,\%, 4.7\,\%, 
12.5\,\%, 20.3\,\%, 29.7\,\%, and 50.0\,\%.
}
\label{fig_area_per_num_vs_time}
\end{figure}

Figure~\ref{fig_area_per_num_vs_time} shows the temporal behavior 
of the area per molecule. It can be seen 
that after 20\,ns the area per molecule has converged even for 
the highest cholesterol concentrations. It is, nevertheless, 
immediately obvious from the data that this type of
MD simulations of bilayer systems should be at least 
of the order of tens of nanoseconds 
to reach equilibrium and surpass the longest characteristic time 
scales for area fluctuations. The first 
20\,ns of the total 100\,ns were therefore considered as 
equilibration, and the last 80\,ns were used for analysis.

The data clearly show the condensing 
effect of cholesterol: the area per molecule decreases with the 
cholesterol content. Further, an increasing cholesterol 
concentration seems to suppress the fluctuations in the 
average area per molecule. The values of the average area per molecule
(see also Fig.~\ref{fig_area_per_num_vs_conc}) 
are in excellent agreement with two recent simulation studies 
on the DPPC\,/\,cholesterol system~\cite[]{Chi02,Hof03}. As for 
experimental results, we are only aware of an accurate measurement 
for the average area per molecule in the case of a pure DPPC 
bilayer~\cite[]{Nag00}. In this case the average area per molecule was 
determined to be $0.64$\,nm$^2$ at 
$T = 323$\,K, in good agreement with $(0.655 \pm 0.005)$\,nm$^2$ 
obtained here. Measurements of the average area per molecule in 
DPPC\,/\,cholesterol monolayers~\cite[]{McC03} show trends 
similar to ours. The exact correspondence between average 
areas per molecule measured for bilayers and monolayers, 
however, is not evident~\cite[]{Nag00}.

\subsection{Ordering of Acyl Chains}

Average areas per molecule are closely related to order 
parameters~\cite[]{Pet99}, which are a measure of the orientational 
order of the phospholipid tails. Order parameters can be 
obtained from deuterium NMR experiments~\cite[]{See74} or 
computer simulations~\cite[]{Tie97}. In united atom simulations 
such as ours, the orientational order can be characterized 
using tensors with elements $S_\mathrm{\alpha \beta}$ such that 
\begin{equation} \label{eq_order_parameter_tensor}
S_\mathrm{\alpha \beta} \equiv \frac{3}{2}\langle \cos \theta_\mathrm{\alpha} 
\cos \theta_\mathrm{\beta} \rangle - \frac{1}{2},
\end{equation}
where $\theta_\mathrm{\alpha}$ is the angle between the molecular 
$\mathrm{\alpha}$ axis 
and the bilayer normal~\cite[]{Tie97}. 
The molecular axes must be defined separately for each segment of 
an acyl chain: usually for the $n^{\mathrm{th}}$ methylene group 
denoted as $C_n$, the z axis points in the $C_{n-1}-C_{n+1}$ direction, 
and $C_{n-1}$, $C_n$, and $C_{n+1}$ span the yz plane. If the motion 
of the segments is assumed to be symmetric about the bilayer normal, the 
experimental deuterium order parameter $S_{CD}$ can be easily acquired: 
\begin{equation} \label{eq_deuterium_order_parameter}
S_\mathrm{CD} = -\frac{1}{2} S_\mathrm{zz}.
\end{equation}
As the two acyl chains {\it sn}--1 and {\it sn}--2 give rise to 
slightly different NMR quadrupole splittings~\cite[]{See74}, it is 
useful to compute the order parameters separately for both chains.

\begin{figure}[h]
\centering
\includegraphics[width=\columnwidth,bb=105 100 425 515,scale=1,clip=true]{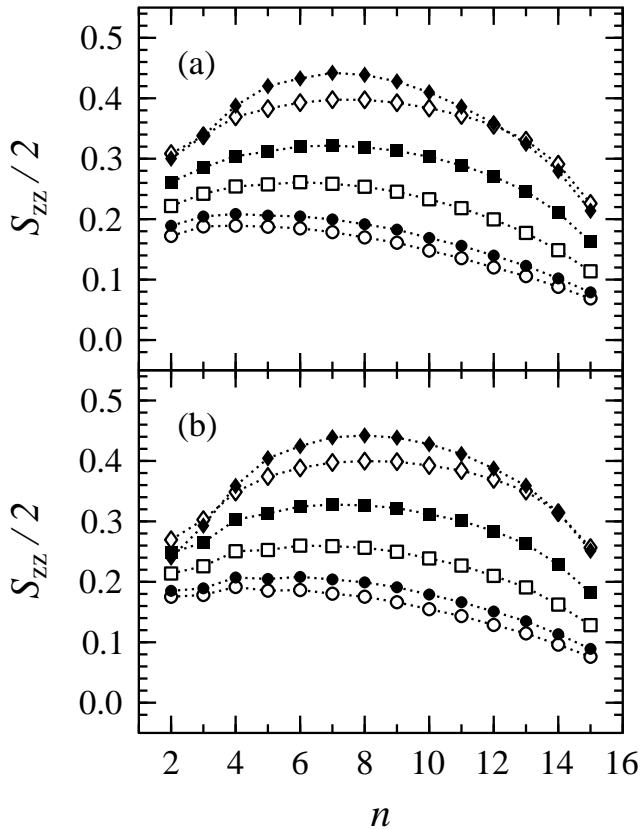}
\caption{
Order parameter profiles for (a) {\it sn}--1 and (b) {\it sn}--2 
tails. The  cholesterol concentrations are: 0.0\,\% ($\circ$), 
4.7\,\% ($\bullet$), 12.5\,\% ($\square$), 20.3\,\% ($\blacksquare$), 
29.7\,\% ($\lozenge$), and 50.0\,\% ($\blacklozenge$), 
and the index $n$ increases towards the center of the bilayer.
}
\label{fig_s_vs_n}
\end{figure}

The order parameter profiles for the {\it sn}--1 and {\it sn}--2 
chains are depicted in Fig.~\ref{fig_s_vs_n}. The ordering effect 
of cholesterol is clearly visible: the order parameters grow 
significantly with an increasing cholesterol content. For 
pure DPPC and low cholesterol concentrations, the order 
parameter profiles show a plateau for small and intermediate 
values of $n$ and decay near the center of the bilayer. When 
the cholesterol content increases, the plateau disappears, 
and there is a clear maximum at intermediate $n$. The ordering 
effect of cholesterol is most pronounced for $n \sim 6-10$ and 
quite modest for segments near the phospholipid headgroups and 
bilayer center. This is due to the position of the cholesterol 
ring system in the bilayer along the bilayer normal~\cite[]{Smo99}: 
the largest ordering occurs for segments at roughly the same depth 
as the ring system. For instance with 30\,\% cholesterol the 
order parameters for $n \sim 6-10$ are increased roughly by 
a factor of two.

Our results for the order parameters are in good agreement 
with other simulation studies~\cite[]{Smo99,Chi02,Hof03}. 
However, as most force fields yield qualitatively similar 
results, and various technical details may influence the detailed
form of the order parameter profile~\cite[]{Pat03a}, it is more 
interesting to make comparisons to experimental findings.

The results for the pure DPPC system are in 
good agreement with experiments~\cite[]{Bro79,Dou95,Pet00a}. 
As for mixtures of DPPC and cholesterol, 
Sankaram and Thompson found that when $50$\,\% of the DPPC 
molecules were substituted by cholesterols in a pure DPPC 
bilayer at $T = 325$\,K, the order parameter for intermediate 
$n$ was increased by a factor of $2.65$~\cite[]{San90b}. 
Similarly, when 30\,\%  of the DMPCs were replaced by 
cholesterols in a pure DMPC bilayer at $T = 308$\,K, the 
order parameter increased by a factor of two. Vist and Davis, 
in turn, observed an increase by a factor of two when 
replacing 24\,\%  of the DPPC molecules by cholesterol 
at $T = 323$\,K~\cite[]{Vis90}. Similar agreement is 
found when our results are compared to other 
experiments~\cite[]{Dou96,Kin86}. In all, our simulations 
agree well with experimental findings. The only detail which 
our, or any other united atom MD simulations cannot reproduce 
is the behavior of the experimental deuterium order parameter 
for {\it sn}--2 at $n = 2$~\cite[]{San90b,See75}.

\subsection{Electron Density Profiles}

Additional information about the structure of the bilayer 
along the normal or z direction can be obtained by computing 
density profiles for the whole system, different molecular 
species, or certain atomic groups of interest. In simulations 
it is possible to calculate atom density, mass density, and 
electron density profiles. These give information on the 
distribution of atoms in the normal direction. Related 
information can be acquired from X-ray and neutron 
diffraction studies. Due to fluctuations, X-ray diffraction 
studies on fully hydrated bilayers in a fluid phase only 
yield total electron density profiles, whose maxima are 
associated with the electron dense phosphate groups~\cite[]{Nag00}. 
The distance between the maxima allows one to estimate the 
distance between the headgroups in the opposite leaflets, 
but does not yield accurate predictions for the hydrocarbon 
thickness or the true phosphate-phosphate distance~\cite[]{Nag96}. 
Additional information, most importantly about the average 
location of various atomic groups, can be gained from neutron 
diffraction studies either with selective deuteration or in 
combination with X-ray diffraction~\cite[]{Nag00}. 

\begin{figure}[h]
\centering
\includegraphics[width=\columnwidth,bb=105 295 425 535,scale=1,clip=true]{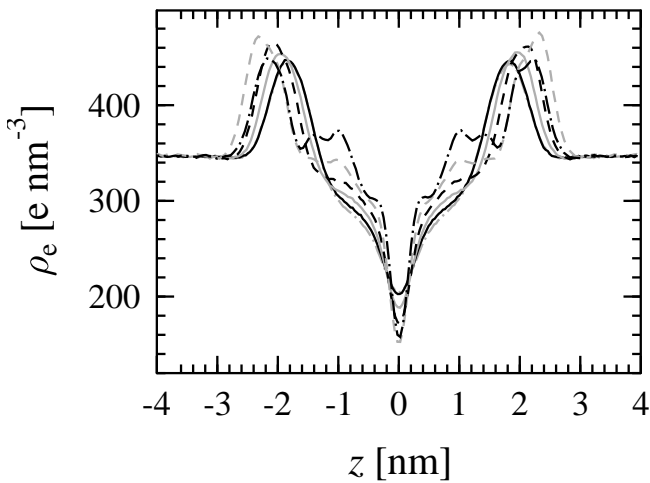}
\caption{
Total electron density profiles as functions of distance z
from bilayer center. The curves correspond to the various 
cholesterol concentrations as follows: 0.0\,\% (dash-dotted grey), 
4.7\,\% (solid black), 12.5\,\% (solid grey), 
20.3\,\% (dashed black), 29.7\,\% (dashed grey), 
and 50.0\,\% (dash-dotted black).
}
\label{fig_eden_tot}
\end{figure}

Figure~\ref{fig_eden_tot} shows the total electron densities 
calculated for the different cholesterol concentrations. The 
density profiles have a characteristic shape reminiscent of 
X-ray diffraction studies, with maxima approximately corresponding 
to the location of the phosphate groups, and a minimum, 
a so-called methyl trough, in the bilayer center, where the 
terminal methyl groups reside. For pure DPPC and low cholesterol 
concentrations, the densities decrease monotonically from the 
maxima to the minimum in the bilayer center. This medium density 
region corresponds to the methylene groups in the DPPC tails. 
When more cholesterol is present, the headgroup-headgroup
distance increases, i.\,e.,~the bilayer gets thicker, and the 
electron density in the bilayer center decreases slightly. In 
addition, the density in the tail region increases, and the 
density profile between the center and the headgroups is no 
longer monotonically decreasing. The elevation is due to the 
fact that the cholesterol ring structure, which resides in the 
phospholipid tail region, has a higher electron density 
than do phospholipid tails.

\begin{figure}[h]
\centering
\includegraphics[width=\columnwidth,bb=5 125 510 580,scale=1,clip=true]{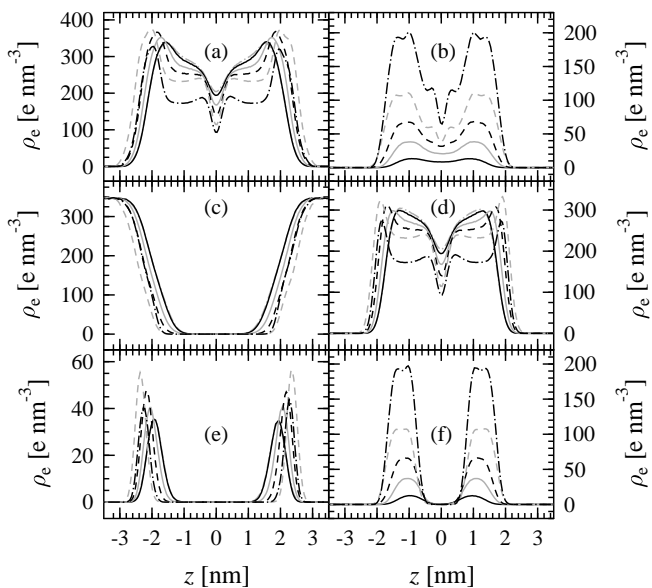}
\caption{
Electron density profiles for molecular species and atomic 
groups: (a) DPPC, (b) cholesterol, (c) water, (d) DPPC tails, 
(e) phosphorus atom (P), and (f) cholesterol ring system. 
The curves correspond to the cholesterol concentrations as 
indicated in Fig.~\ref{fig_eden_tot}.
}
\label{fig_eden}
\end{figure}

To gain more insight into the structure of the bilayer, 
we can investigate the electron densities for DPPC molecules, 
cholesterols, water molecules, phospholipid tails, phosphorus 
atoms, and cholesterol rings, portrayed in Fig.~\ref{fig_eden}.
All density profiles are consistent with a thickening of 
the bilayer with an increasing amount of cholesterol: the 
molecules and their constituent atomic groups are pushed 
towards the water phase. Still, it is clear that for all 
cholesterol concentrations, DPPC molecules largely stay 
within a distance of 3\,nm from the center, while 
cholesterols and DPPC tails can be found within about 2\,nm. 
We can conclude that cholesterol is located in the hydrophobic 
interior of the bilayer. The penetration of water into the 
bilayer becomes more difficult with increasing amounts of 
cholesterol: this reflects both the thickening of the bilayer, 
and the increasing densities in the headgroup region. The 
lipid\,/\,water interface also seems to become steeper. The 
electron density of DPPC in the hydrophobic tail region 
decreases with the cholesterol content, which is compensated 
by an increasing cholesterol electron density. By comparing 
the electron densities for cholesterol and cholesterol ring 
systems, we can conclude that only the short acyl chain of 
cholesterol can approach the bilayer center.

Both the total electron density profile and the densities 
for molecular species and atomic groups can be compared to 
previous simulations. Here we will concentrate on simulations 
on DPPC with cholesterol at $T = 323$\,K~\cite[]{Hof03,Smo99,Tu98}. 
In all simulation studies, the peaks that indicate the location 
of headgroups for pure DPPC are located approximately at the 
same distance from the bilayer center. With 10\,--\,12.5\,\% 
cholesterol only minor changes in the total densities can be 
observed. Except in the case of Tu et al., increasing amounts 
of cholesterol lead to a larger bilayer thickness and a slightly 
decreased total density in the bilayer center. All studies 
show an increased density in the phospholipid tail region.  
By investigating the DPPC or water densities, one may also 
note that all studies clearly indicate that the lipid\,/\,water 
interface becomes more abrupt. Our findings for the distribution  
of phosphorus atoms agree well with those of Smondyrev et al.: 
when the cholesterol content increases, the peaks are narrowed 
and shifted towards the water phase. In all, density profiles 
computed using slightly different force fields are, for the 
most part, consistent with each other.

Our results are also consistent with diffraction experiments 
on DPPC and DMPC bilayers. Nagle et al.~have determined the 
structure of a fully hydrated pure DPPC bilayer in the liquid 
disordered phase using X-ray diffraction~\cite[]{Nag96}. The form 
of the density profile from our simulations of pure DPPC closely 
resembles Nagle's electron density profile for pure DPPC at 
$T = 323$\,K. The head-head distance obtained from Nagle's 
experiment and that determined from our density profiles also 
are in good agreement. As for the influence of cholesterol, 
McIntosh has published X-ray diffraction experiments on model 
membranes containing cholesterol and phospholipids with saturated 
tails containing 12\,--\,18 carbons~\cite[]{McI78}. His
DLPC\,/\,cholesterol systems in the fluid phase 
behave in a qualitatively similar way as do our DPPC\,/\,cholesterol 
bilayers. By comparing the electron densities from systems 
with different phospholipids and cholesterol to the 
densities from pure phospholipid bilayers, McIntosh also 
establishes the location of the cholesterol ring structure 
in the bilayer. Our studies support his view. In addition, 
there are more recent neutron diffraction studies of 
DMPC\,/\,cholesterol bilayers. The studies by Douliez et al. 
and L\'eonard et al. clearly show that substituting 30\,\% 
of the phospholipids by cholesterol in a pure DMPC bilayer 
in the liquid disordered phase increases the bilayer 
thickness~\cite[]{Dou96,Leo01}. L\'eonard et al.~have also investigated the 
location of cholesterol in the bilayer, and concluded that 
cholesterol is located well within the hydrophobic core. 
Although DPPC has longer hydrocarbon tails than DMPC, the 
cholesterol ring structure should be located in the same 
region of the bilayer~\cite[]{McI78}. Our simulations indicate 
that cholesterol is indeed situated in the non-polar region, 
as is the case in Douliez's and McIntosh's experiments.

\subsection{Radial Distribution Functions}

Together, the above results ascertain that our model correctly 
describes the behavior of the dimensions of the bilayer and 
the ordering of the non-polar phospholipid tails as functions 
of the cholesterol content. Further, the structure of our 
DPPC\,/\,cholesterol bilayer in the normal direction is 
consistent with results from previous computations and 
experiments. This is very satisfactory, but in addition, we 
need to ensure that our bilayers truly are in the fluid 
state, i.\,e.,~that there is no translational long-range order. 
This can be ascertained by examining the radial distribution 
functions (RDFs) for e.\,g.~phosphorus and nitrogen atoms 
in the DPPC headgroups. For instance the N\,--\,N 
radial distribution functions calculated in two dimensions
for various cholesterol concentrations have large 
nearest-neighbor peaks at $r \sim 0.82$\,nm and show essentially 
no structure beyond $r = 1.5$\,nm (data not shown).
Additional calculations for other pairs of atoms and for 
the center of mass positions of the DPPC and cholesterol molecules 
lead to a similar conclusion, i.\,e.,~there is no lateral long-range 
structure. Hence, we can be confident that our bilayers are either 
in the liquid disordered or liquid ordered phase, as they should. 
With this, we consider our model to be valid.

\subsection{Estimating Average Areas per Molecule in Multi-Component Bilayers
\label{sec_area_per_molecule}}

The average area per molecule, which is obtained by dividing the 
total area of the bilayer by the total number of molecules, 
is a well-defined concept in one-component lipid bilayers. 
It includes both area actually occupied by a lipid, 
so-called close-packed area, and some free area. 
A similar quantity can be defined for 
multi-component bilayers. It is a useful quantity
when simulation results are compared to experiments.
Its interpretation, however, is less clear: different  
lipids and sterols could occupy significantly different amounts of area.
Hence, it would be desirable to be able to estimate the 
average area occupied by each molecular species 
present in the bilayer.

\begin{figure}[h]
\centering
\includegraphics[width=\columnwidth,bb=105 295 425 535,scale=1,clip=true]{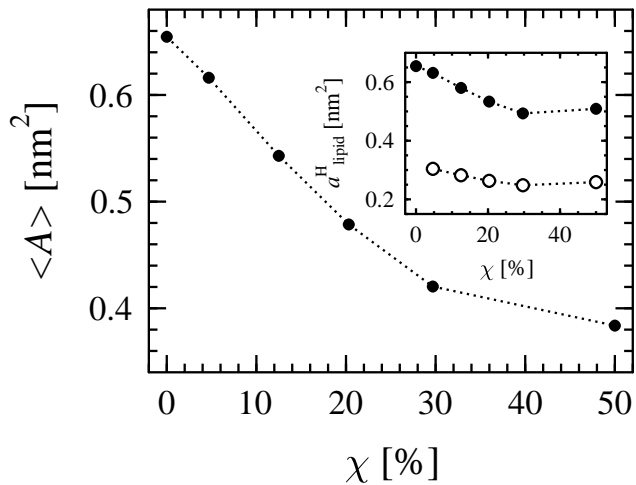}
\caption{
Average area per molecule as function of cholesterol 
concentration. The inset shows the average areas per DPPC 
($\bullet$) and cholesterol ($\circ$) computed as in 
a recent simulation study by Hofs\"a{\ss} et al. 
The errors are smaller than the markers.
}
\label{fig_area_per_num_vs_conc}
\end{figure}

The average area per molecule $\langle A \,\rangle$ as a 
function of cholesterol concentration $\chi$ is portrayed in 
Fig.~\ref{fig_area_per_num_vs_conc}. As mentioned in 
Section~\ref{sec_eq}, it is evident that $\langle A \,\rangle$ 
decreases with the cholesterol content, and that the results 
agree well with previous simulation studies~\cite[]{Chi02,Hof03}.

We would not, however, like Chiu et al., dare to hazard a guess 
that $\langle A \,\rangle$ decreases linearly with $\chi$ and 
use this assumption to compute the average areas per phospholipid 
and cholesterol. It is not obvious, in the first place, that the 
average area per cholesterol or DPPC is independent of cholesterol 
content, as these authors seem to imply.

Another way to divide the total area between DPPC and 
cholesterol molecules has also been suggested~\cite[]{Hof03}. 
By computing the total area and volume of the simulation 
box as functions of the cholesterol content and making 
a number of assumptions, one can arrive at estimates for 
the average areas occupied by DPPC and cholesterol molecules. 
In this case, an important assumption is that the average 
volume of a cholesterol molecule can be, for all concentrations, 
taken to be the volume occupied by a cholesterol molecule in 
a cholesterol crystal. Further, it is assumed that all space 
is occupied by DPPC, cholesterol, or water, i.\,e.,~that there 
is no free volume or area. The average areas per DPPC and 
cholesterol, $a^\mathrm{H}_\mathrm{DPPC}$ and 
$a^\mathrm{H}_\mathrm{chol}$, 
obtained along these lines from our data, are shown in the 
inset of Fig.~\ref{fig_area_per_num_vs_conc}. These closely 
resemble the corresponding results by Hofs\"a{\ss} et al.

\begin{figure}[h]
\centering
\includegraphics[width=\columnwidth,bb=105 295 425 535,scale=1,clip=true]{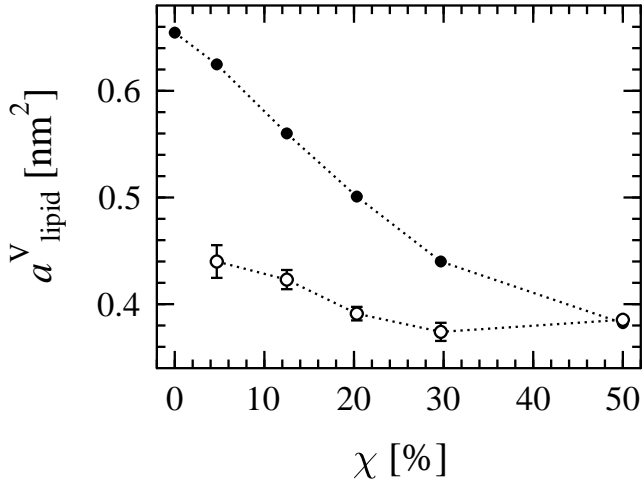}
\caption{
Areas per DPPC ($\bullet$) and cholesterol ($\circ$) 
computed using Voronoi tessellation. 
The errors for DPPC are smaller than the markers.
}
\label{fig_area_vor}
\end{figure}

A yet further method of distributing the area among the 
molecular species in a bilayer is to apply Voronoi analysis 
in two dimensions~\cite[]{Shi98a,Pat03a}. In Voronoi tessellation 
for a bilayer, the center of mass (CM) coordinates of the 
molecules comprising the bilayer are projected onto the xy 
plane. An arbitrary point in this plane is considered to 
belong to a particular Voronoi cell, if it is closer to the 
CM position associated with that cell than to any other one. 
In this way one can calculate the total area associated with 
the CM positions of e.\,g.~the DPPC molecules and then scale 
this quantity by the number of DPPC molecules in a monolayer. 
The resulting average areas per DPPC and cholesterol, 
$a^\mathrm{V}_\mathrm{DPPC}$ and $a^\mathrm{V}_\mathrm{Chol}$, 
as functions of the cholesterol content, 
are depicted in Fig.~\ref{fig_area_vor}. 

These values for the areas per DPPC and cholesterol do 
differ from those reported by Hofs\"a{\ss} et al. This is 
due to the fact that the assumptions inherent to the respective 
methods lead to different ways of distributing the free area 
in the bilayer. Further, basic Voronoi analysis does not, in 
any way, allow one to take into account the close-packed sizes 
of the molecules. It may well be that the area, 
which from the point of view of the Voronoi analysis belongs 
to cholesterol, as a matter of fact, would be covered by 
projected coordinates of atoms from a DPPC molecule.

\subsection{Slicing Membranes} 

We are now confronted by fundamental questions relevant to both
one- and multi-component bilayer systems. How can we find 
estimates for the average close-packed 
cross-sectional areas for the molecular species present in 
a one-component or composite bilayer? Further, how can 
we estimate the average amount of free area in a membrane?

Our approach to answer these questions bears a certain 
resemblance to tomography. We map each configuration on a 
number of rectangular three-dimensional grids as follows. 
If a grid point lies within the van der Waals radius of an atom 
belonging to a DPPC molecule, this point is considered occupied, 
and otherwise empty, on a grid keeping account of DPPC molecules. 
Grid points within van der Waals radiae of atoms belonging to 
cholesterol, in turn, will be occupied on a grid characterizing 
the cholesterol molecules. Finally, a grid for water molecules 
is constructed analogously. In the xy plane the grids have 
$100 \times 100$ elements. Because the system size fluctuates 
weakly, the size of an element will vary slightly from 
configuration to configuration. In the z direction, on the 
other hand, the size of the elements has been fixed to 0.1\,nm, 
and we only consider grid points within 3\,nm from the bilayer 
center.

\begin{figure}[h]
\centering
\includegraphics[width=\columnwidth,bb=30 135 480 635,scale=1,clip=true]{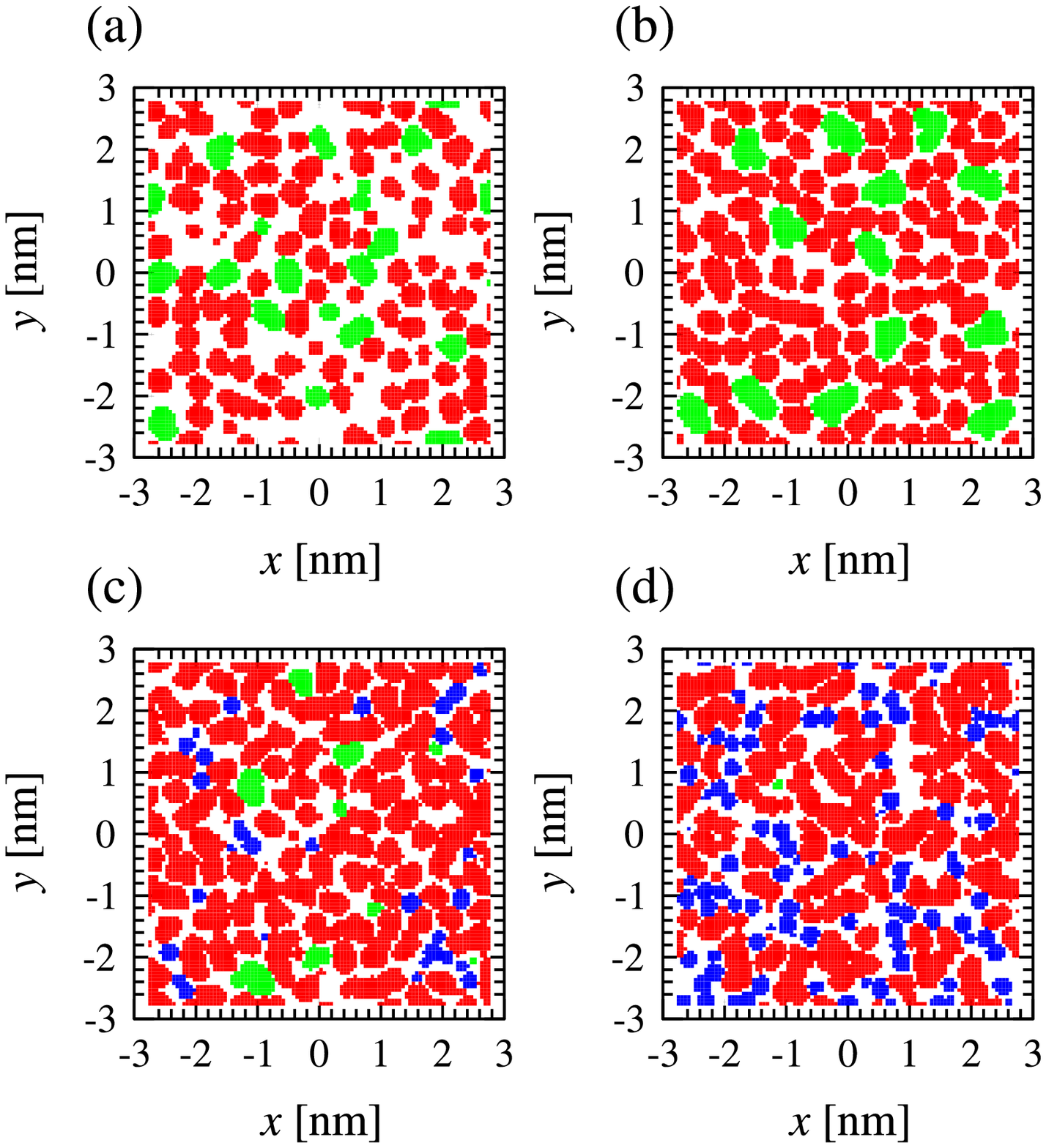}
\caption{
Cross-sections of bilayer with 20\,\% cholesterol at 100\,ns. 
DPPC grid elements have been coloured red, cholesterol is green, 
and water blue. The remaining area, i.\,e.,~the free area, is white. 
The panels correspond to slices at different distances $z$ from 
the bilayer center: (a) bilayer center, (b) $z \sim 1$\,nm, 
(c) $z \sim 1.7$\,nm, (d) $z \sim 2$\,nm.
}
\label{fig_cut}
\end{figure}

The grids can be used to view given slices of the bilayers: 
they show cross-sections of DPPC, cholesterol, and water 
molecules, as well as patches of free area. Pictures of slices can 
be illustrative as such, and Fig.~\ref{fig_cut} contains 
a selection of such slices for the case of 
20\,\% cholesterol. From Figs.~\ref{fig_eden} 
and \ref{fig_cut}(a) we can conclude that there is quite 
large amounts of free area in the bilayer center, and that 
cholesterol tails from a given monolayer extend to the opposite 
monolayer. Panel (b) in Fig.~\ref{fig_cut} portrays the region 
where DPPC tails and cholesterol ring structures should, according 
to Fig.~\ref{fig_eden}, dominate. DPPC tails can be recognized 
as circular red structures, and the green formations are 
cross-sections of cholesterol ring structures. 
Part (c) is a cross-section of the bilayer at a distance 
$z \sim 1.7$\,nm from the bilayer center. Some cholesterol 
is still present in this slice, and there are also small 
amounts of water. The amount of free area is significantly 
smaller than in the bilayer center. Panel (d) finally shows 
a cross section at $z \sim 2$\,nm: there are DPPC headgroups, 
substantial amounts of water, and very little cholesterol.

\begin{figure}[h]
\centering
\includegraphics[width=\columnwidth,bb=105 295 425 535,scale=1,clip=true]{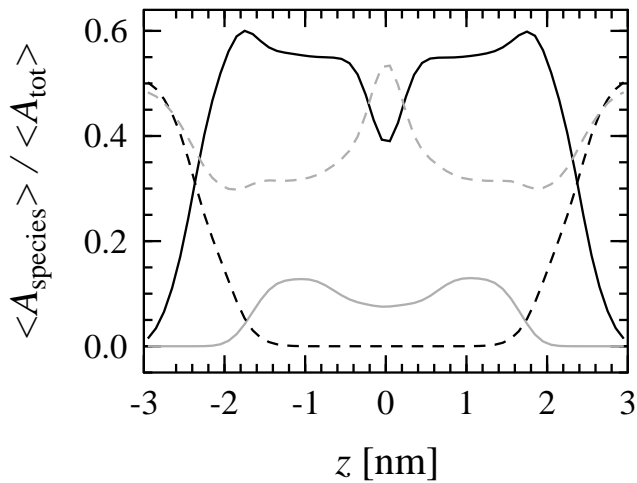}
\label{fig_all_areas}
\caption{
Area profiles for bilayer with 20\,\% cholesterol scaled by total 
bilayer area: DPPC area profile 
$\langle A_\mathrm{DPPC}(z) \rangle / \langle A_\mathrm{tot} \rangle$ 
(solid black), cholesterol area profile 
$\langle A_\mathrm{chol}(z) \rangle / \langle A_\mathrm{tot} \rangle$ 
(solid grey), water area profile 
$\langle A_\mathrm{water}(z) \rangle / \langle A_\mathrm{tot} \rangle$ 
(dashed black), free area profile 
$\langle A_\mathrm{free}(z) \rangle / \langle A_\mathrm{tot} \rangle$ 
(dashed grey). The errors of the scaled areas are of the order of 
a few per cent.
}
\end{figure}

From the grids constructed for DPPC, cholesterol, and water, 
we can compute total area profiles for the various 
molecular species, that is, average total areas 
occupied by the molecules as functions of the distance from the 
bilayer center. In addition, we can calculate free area profiles, 
i.\,e.,~the amount of free area as a function of the distance 
from the bilayer center. In practice, this is achieved by 
traversing the grids slice by slice and augmenting the 
various area profiles. If a grid element in a certain slice 
at a distance $z$ from the bilayer center is occupied 
in, say, the DPPC grid, but not in 
the cholesterol or water grids, we increment the total area 
of DPPC in that slice, $A_\mathrm{DPPC}(z)$, by an area 
corresponding to a grid element. If, on the other hand, 
a grid point at a distance $z$ from the center is occupied 
by neither DPPC nor cholesterol, nor water, the total free 
area $A_\mathrm{free}(z)$ in the slice in question is incremented. 
In the end we average over the total area profiles constructed 
separately for each configuration. This procedure leads to 
a definition of free area which is similar in nature to the concept 
of empty free volume introduced by Marrink et al.~to characterize 
a pure DPPC bilayer~\cite[]{Mar96}. 
Figure~\ref{fig_all_areas} exemplifies the computation of the 
various area profiles for a bilayer with 20\,\% cholesterol.

\subsection{Close-Packed Areas for DPPC and Cholesterol}

To gain understanding of the effect of cholesterol
on the properties of phospholipid bilayers, 
we first concentrate on the behavior of the cross-sectional 
area profiles for DPPC and cholesterol. Hence, we need to 
know both the total areas of DPPC and cholesterol and 
the average numbers of respective molecules as functions of the 
distance from the bilayer center. The total areas occupied by 
DPPC and cholesterol molecules, denoted by 
$\langle A_\mathrm{DPPC}(z) \rangle$ and 
$\langle A_\mathrm{chol}(z) \rangle$, 
for the different cholesterol concentrations are computed 
in the manner described above.

\begin{figure}[h]
\centering
\includegraphics[width=\columnwidth,bb=105 100 425 515,scale=1,clip=true]{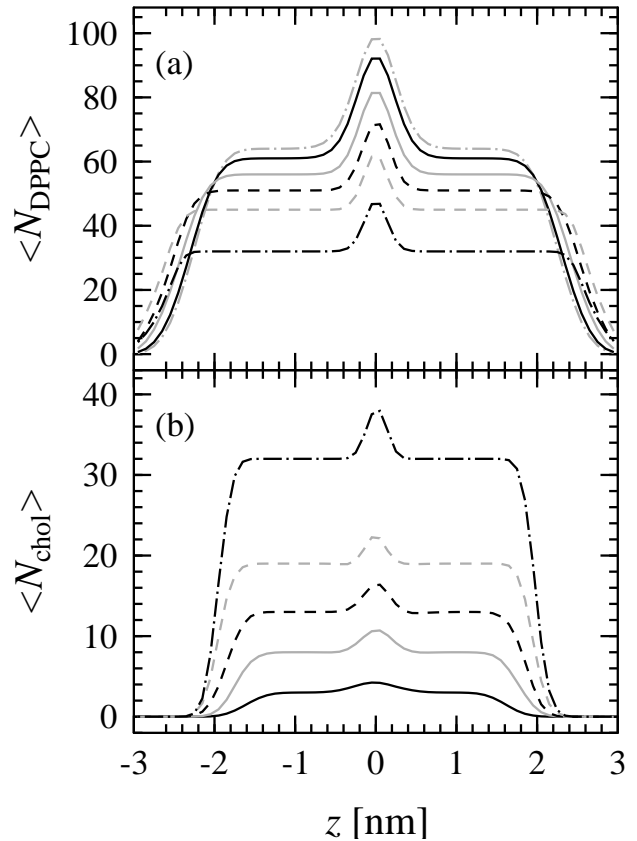}
\caption{
Numbers of (a) DPPC and (b) cholesterol molecules as functions of
distance from bilayer center. The curves correspond to the
cholesterol concentrations as indicated in Fig.~\ref{fig_eden_tot}.
The errors are of the order less than 1\,\%.
}
\label{fig_num}
\end{figure}

To find the average numbers of DPPC and cholesterol 
molecules in each slice, we locate the maximum and minimum $z$ 
coordinates of each molecule with respect to the bilayer center, 
taking into account the finite size of the constituent atoms. 
The molecule is considered to be present in all the slices 
between these points. By averaging over all molecules of 
a certain species and over all configurations, we arrive at 
the average numbers of DPPC molecules and cholesterols as 
functions of the distance from the bilayer center, denoted by 
$\langle N_\mathrm{DPPC}(z) \rangle$ and 
$\langle N_\mathrm{Chol}(z) \rangle$, shown in Fig.~\ref{fig_num}. 
Perhaps the most notable feature in Fig.~\ref{fig_num} is that all 
curves peak in the bilayer center. This is due to so-called 
interdigitation: a substantial part of both DPPC and cholesterol 
molecules extend to the opposite monolayer. On both sides
of the peak, there are broad plateaus, which reflect the amount
of molecules of a certain species in a monolayer. Eventually, at 
about $3$\,nm from the bilayer center for DPPC and $2$\,nm for 
cholesterol, the curves decay to zero.

There seem to be two effects that together
contribute to the thickening of the bilayer, both 
visible in Fig.~\ref{fig_num}. First, the DPPC
molecules are extended. Cholesterol 
molecules, on the other hand, are not significantly elongated. 
These observations are quite plausible, as the 
presence of cholesterol leads to a smaller
amount of gauche defects in the acyl chains of the DPPC 
molecules~\cite[]{Hof03}. Further, the
tilt of the DPPC molecules with respect to the bilayer normal 
decreases~\cite{Rog01a}. Cholesterol, on the other hand, with its rigid
ring structure and short tail, does not undergo such significant extension.

The second effect partially responsible for the thickening 
is that with a larger cholesterol content $\chi$,
a smaller amount of DPPC and cholesterol molecules extend to 
the opposite monolayer. Further, the ones that protrude do 
not penetrate quite as deep into the opposite leaflet as they 
do at low cholesterol concentrations. In a 
pure DPPC bilayer, $53$\,\% of the DPPC molecules protrude to the
opposite monolayer, while at $29.7$\,\% cholesterol the corresponding
figure is $40$\,\%. The effect is stronger for cholesterol:
At $4.7$\,\% and $29.7$\,\% concentrations, respectively,
$41$\,\% and $17$\,\% of the molecules extend to the opposite bilayer.
As the cholesterol hydroxyl is thought to be anchored to the
DPPC headgroup via direct hydrogen bonding or through water 
bridges~\cite[]{Chi02,Pas00}, this effect may be coupled to 
the elongation of the DPPC molecules.

\begin{figure}[h]
\centering
\includegraphics[width=\columnwidth,bb=105 100 425 515,scale=1,clip=true]{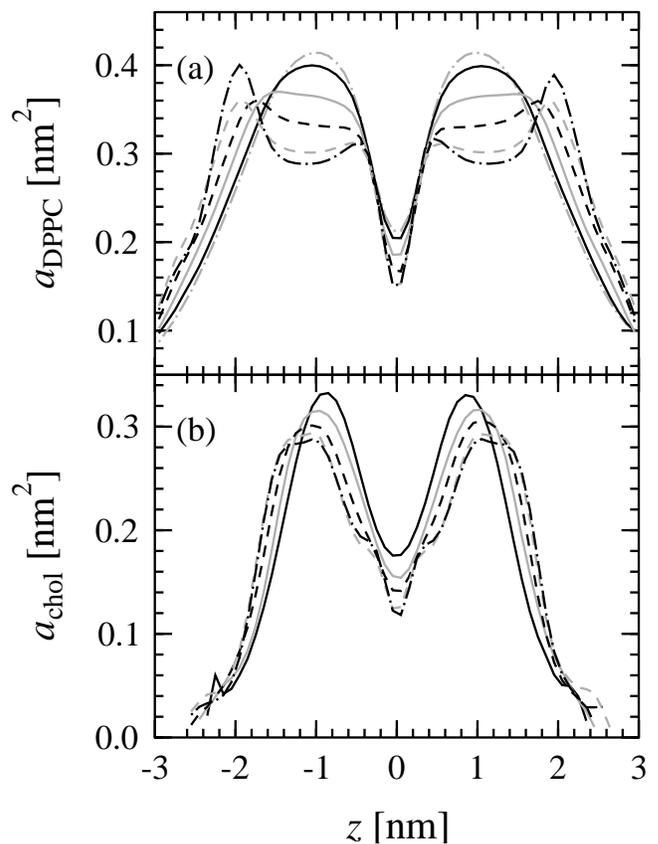}
\caption{
Cross-sectional close-packed areas for (a) DPPC and (b) cholesterol 
molecules as functions 
of distance from bilayer center. The curves correspond to 
the cholesterol concentrations as indicated in Fig.~\ref{fig_eden_tot}. 
The errors are of the order of a few percent.
In the water phase, the relative errors for 
$a_\mathrm{chol}$ are somewhat larger.
}
\label{fig_area_per_num}
\end{figure}

Equipped with the total areas occupied by the molecular species
together with the average numbers of these 
molecules as functions of distance from the bilayer center, we 
can now compute the average cross-sectional areas for DPPC 
and cholesterol across a membrane, 
$a_\mathrm{DPPC}(z) \equiv \langle A_\mathrm{DPPC}(z) \rangle / 
\langle N_\mathrm{DPPC}(z) \rangle$
and 
$a_\mathrm{chol}(z) \equiv \langle A_\mathrm{chol}(z) \rangle / 
\langle N_\mathrm{chol}(z) \rangle$, shown in 
Fig.~\ref{fig_area_per_num}.

It should not come as a surprise that the cross-sectional 
close-packed area occupied by a DPPC or cholesterol molecule is
not constant along the bilayer normal. In the 
case of DPPC, there are significant changes in the form 
of $a_\mathrm{DPPC}(z)$ when the cholesterol content is 
increased. A maximum located at approximately 1\,nm from the 
bilayer center for a pure DPPC bilayer becomes at 
intermediate cholesterol concentrations 
a plateau at $0.5 - 1.5$\,nm from the center, and finally 
with $29.7$\,\% cholesterol in the bilayer develops into 
two small maxima at 0.5\,nm and 2\,nm with a shallow minimum 
in between.

These changes in $a_\mathrm{DPPC}(z)$ are for the most 
part due to the behavior of the phospholipid tails: they 
occur in regions where the tail densities are high and where 
there are little or no headgroups. This can be deduced by
comparing the electron densities for DPPC molecules and DPPC 
tails in Figs.~\ref{fig_eden}(a) and (d). In addition, we should 
note that the electron density profiles for the DPPC tails and the  
$a_\mathrm{DPPC}(z)$ curves have many features common.
This allows us to, at least partially, interpret the 
behavior of $a_\mathrm{DPPC}(z)$ from the point 
of view of ordering. The most substantial ordering effect 
with large amounts of cholesterol present in the bilayer 
occurs for carbons in the middle of the tail, see 
Fig.~\ref{fig_s_vs_n}. Close to the headgroups and in the bilayer 
center the ordering effects of cholesterol are more modest.
As increased order correlates with a decreasing area occupied
by the tails, one expects that with an increased cholesterol 
content the cross-sectional area per DPPC approximately at 
a distance 1\,nm from the bilayer should decrease. Our 
findings are consistent with this picture.

This is not to say, however, that there would exist a simple way 
of mapping $a_\mathrm{DPPC}(z)$ with order parameter profiles
(see also Section~\ref{sect_ordering_area}). The maximum that 
develops at $z \sim 2$\,nm, e.\,g., is a result of contributions 
from both tails and headgroups. From separate cross-sectional 
area profiles for the two tails and the headgroup of a DPPC 
molecule (data not shown), we found that when the cholesterol 
concentration increases, the cross-sectional area occupied by 
the tail portion of a DPPC molecule decreases as a consequence 
of ordering, while the area occupied by the headgroup seems to 
be increasing slightly (data not shown). The maximum at 
$z \sim 2$\,nm at intermediate and high cholesterol 
concentrations is hence related to the interplay of the 
decreasing tail contribution with a plateau centered at 
$z \sim 1$\,nm and the slightly increasing head contribution 
that peaks at $z \sim 2$\,nm.

In the case of cholesterol the cross-sectional close-packed area 
of a molecule is changed only weakly when the cholesterol concentration
$\chi$ is increased. The slight decrease with an increasing $\chi$ can be 
explained by the tilt of the cholesterol molecules.
At high concentrations, almost all cholesterols
are oriented nearly parallel to the bilayer normal 
(data not shown). At low concentrations, on the other hand, the 
distribution of the angle between the bilayer normal and the 
ring structure becomes more broad and flat, i.\,e.,~the molecules
are more tilted with respect to the bilayer normal. Hence
the cross-sections appear larger at low concentrations.

The general form of $a_\mathrm{chol}(z)$ is compatible with our 
idea of the structure of the cholesterol molecule: narrow in the 
bilayer center where the small cholesterol tails reside and broad 
where the ring structure is located. It also reflects the thickening 
of the bilayer, as the maxima associated with the ring structures 
are pushed towards the water phase when more cholesterol is 
present. This picture, overall, supports the common belief 
that the average area per cholesterol in a phospholipid 
bilayer is largely unaltered by the amount of cholesterol 
in the bilayer.

Our results for $a_\mathrm{chol}(z)$ can be compared to the outcome 
of an old experiment~\cite[]{Rot72}, where a model of cholesterol made 
of plastic was immersed in a tube filled with water. This experiment 
resulted in a steric profile for cholesterol, i.\,e.,~a profile 
of the cross-sectional area occupied by cholesterol. This 
steric profile and our $a_\mathrm{chol}(z)$, especially at 
high cholesterol concentrations, bear a surprisingly good 
resemblance to each other. The steric profile measured by
Rothman and Engelman displays a plateau where the cholesterol 
rings are located, with cross-sectional areas of the order of 
0.25\,nm$^2$. In the region where the cholesterol tail is 
located, they report a small maximum: here the cross-sectional 
areas are of the order of 0.15\,nm$^2$.

It is clearly difficult to describe the close-packed area of a
DPPC or cholesterol molecule by a single number. Of course, we 
could attempt to define the close-packed area of e.\,g.~a DPPC 
molecule in a given DPPC\,/\,cholesterol bilayer as the maximum 
of the relevant $a_\mathrm{DPPC}(z)$ profile, but this would 
not give accurate information about the packing of DPPC and 
cholesterol molecules in a composite bilayer. Despite this, 
we may note that the maximum values are useful at least when 
assessing the plausibility of the close-packed area profiles 
for DPPC and cholesterol molecules. 

In the case of DPPC the maxima assume values between 
0.36\,nm$^2$ and 0.42\,nm$^2$. These values can be 
compared to the average area per molecule in a pure DPPC bilayer 
in the gel state, where the contribution of the free area to the 
total area assigned to a phospholipid molecule is expected to 
be rather minor. Experiments have yielded an area per molecule 
of approximately 0.48\,nm$^2$~\cite[]{Nag00}, and MD simulations 
suggest that $\langle A \, \rangle =  0.46$\,nm$^2$~\cite[]{Ven00}. 
An exact comparison is not meaningful, since DPPC\,/\,cholesterol 
mixtures, especially with high cholesterol concentrations, have 
structures quite different from a pure DPPC bilayer in the gel 
state. However, the comparison shows that the magnitude of the 
close-packed areas for DPPC molecules is rational.

In a similar fashion, the maxima of the $a_\mathrm{chol}(z)$ 
profiles can be compared to values extracted from experiments on 
cholesterol crystals. The maxima found in this study decrease 
monotonically from 0.33\,nm$^2$ to 0.29\,nm$^2$ when the 
cholesterol concentration changes from $4.7$\,\% to $29.7$\,\%.
In a cholesterol crystal, the area per cholesterol, which in 
this case contains both occupied and free area, has been reported 
to be 0.38\,nm$^2$~\cite[and references therein]{Cra79,Chi02,Hof03}.

\subsection{Free Area}

\begin{figure}[h]
\centering
\includegraphics[width=\columnwidth,bb=105 295 425 535,scale=1,clip=true]{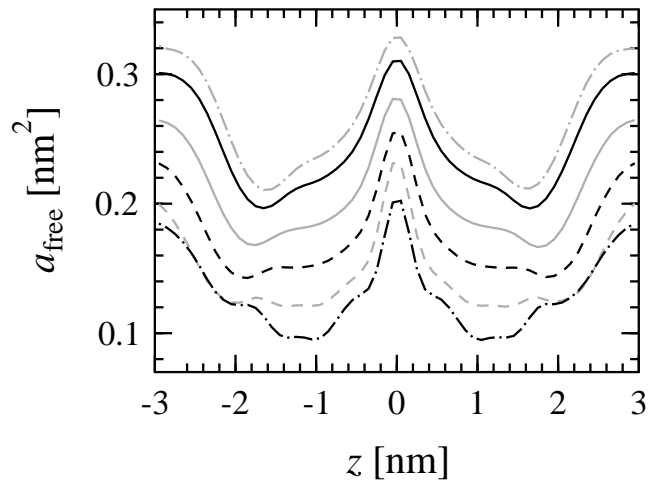}
\caption{
Free areas per molecule as functions of distance from bilayer 
center for different cholesterol concentrations. The curves 
correspond to the cholesterol concentrations as indicated 
in Fig.~\ref{fig_eden_tot}.
}
\label{fig_area_free}
\end{figure}

We now turn our attention to the behavior of free area 
profiles for bilayers with different amounts of cholesterol. 
In Fig.~\ref{fig_area_free}, we show the average amount 
of free area per molecule, 
i.\,e.,~$a_\mathrm{free} \equiv \langle A_\mathrm{free} \rangle / N$, 
where $N$ is the total number of molecules---both phospholipids and 
cholesterol---in a monolayer. The figure clearly shows that the 
amount of free area per molecule decreases in all regions of 
the bilayer, i.\,e.,~for all values of $z$, with an increasing 
cholesterol content. Compared to the case of pure DPPC, 4.7\,\% 
cholesterol in the bilayer leads to a free area per molecule 
reduced by approximately 7\,\% in all regions of the bilayer. 
With 12.5\,\%, 20.3\,\%, and 29.7\,\% cholesterol in the bilayer, 
the free area per molecule is decreased by 20\,\%, 35\,\%, and 
45\,\%. One may note that the behavior 
of the total area of the bilayer cannot be explained by the 
reduced free area only. The occupied area, i.\,e.,~the area 
taken up by DPPC, cholesterol, or water molecules, also decreases 
with more cholesterol. For instance when 30\,\% of the DPPC 
molecules are substituted by cholesterol, the amount 
of occupied area decreases by approximately 30\,\%.

Figure~\ref{fig_area_free} also demonstrates that an increasing 
cholesterol content in a bilayer implies that the form of the 
free area profile is altered. Nevertheless, the different 
curves corresponding to the various cholesterol concentrations 
have certain features in common: the free area profiles all have 
a maximum in the bilayer center, and there is more free area per 
molecule in the water phase than in the tail and headgroup regions. 
For pure DPPC and low cholesterol concentrations, we observe 
a minimum of free area per molecule at $z \sim 1.7$\,nm. For large 
cholesterol concentrations the minimum is still present, but 
due to the thickening of the bilayer it is pushed towards larger 
$z$: e.\,g.~for 30\,\% it can be found at $z \sim 2$\,nm. This 
minimum can, for all cholesterol concentrations, be associated 
with a peak in the density profile of DPPC molecules located 
slightly behind the headgroups in a region where both tails 
and headgroups are present. The density of water in this region is 
already substantial, while there is very little cholesterol.
When the cholesterol concentration is increased, 
also another flat, plateau-like minimum starts 
to develop between the bilayer center and the minimum associated 
with the maximum in the DPPC density, i.\,e.~at $z \sim 1 - 2$\,nm. 
The plateau is almost constant through the tail and headgroup 
regions. It has counterparts in the area profiles of DPPC and 
cholesterol: the cross-sectional DPPC area displays here a flat 
minimum and the cholesterol area a broad maximum 
(see Fig.~\ref{fig_area_per_num}). We can thus conclude
that the changes in the form of the free area profile are intimately
related to modifications in the packing of the molecules in the 
bilayer.

It is evident that the free area profiles are related to the 
relocation and diffusion of solutes inside membranes. MD 
simulations suggest that solutes such 
as ubiquinone~\cite[]{Sod01} and benzene~\cite[]{Bas93} are 
preferentially located in the hydrophobic core region of a 
membrane. Also, it is known that certain non-polar probe 
molecules, e.\,g.~diphenylhexatriene, prefer the bilayer 
center to the lipid\,/\,water interface~\cite[]{Len93}. These 
observations are in accord with our suggestion that the free 
area is largest in the bilayer center.

There are two other simulation studies where quantities 
similar in nature to our free area profile have been calculated 
for DPPC or DPPC\,/\,cholesterol bilayers. Marrink et al.~have 
calculated a so-called empty free volume profile for pure 
DPPC~\cite[]{Mar96}. This should give essentially the same 
information about the amount of average free area in 
a given cross-section of the bilayer as does our free area 
profile for pure DPPC. Our profile does indeed show the 
same general features as Marrink's: a maximum in the bilayer 
center and minima near the headgroup region. Tu et al.~have 
also looked at the influence of 12.5\,\% cholesterol on 
a so-called empty free volume fraction~\cite[]{Tu98}, which 
is equivalent to our total free area scaled by the total 
area of the bilayer. If we compare such scaled free areas 
(data not shown) to Tu's data, we see that the scaled 
profiles have many features in common. 
One difference is that the bilayer thickening is not 
visible in Tu's results, while it can be clearly distinguished 
from ours. The thickening has also been verified experimentally. 
Further, there are some differences in the detailed form of 
the profiles in the bilayer interior, i.\,e.,~the location of the minima 
are slightly different. As Tu et al.~point out, the differences are 
probably due to different computational models.

\subsection{Lateral Diffusion and Free Area \label{sec_diff}}

We have seen that an increasing cholesterol 
concentration reduces the amount of free area per molecule
in the bilayer and simultaneously alters the packing 
of the molecules. On the other hand, it is well
known from experiments that lateral diffusion of 
both DPPC and cholesterol molecules is affected by changes 
in the cholesterol content~\cite[]{Alm92,Fil03,Kon92}. 
It is reasonable to expect that these properties of the 
bilayer and the observed modifications in them with the 
cholesterol concentration are related. Free volume 
theory is a simple but appealing model for explaining such 
dependencies.

Free volume theory was originally 
developed for describing the transport properties of glass-forming 
fluids~\cite[]{Coh59,Mac65,Tur61,Tur70}. It was subsequently adapted 
to modeling two-dimensional diffusion~\cite[]{Gal79,Mac82,Vaz85,Alm92} 
and is usually in this context dubbed free area theory. Free 
area theory, a two-dimensional mean-field model for diffusion, 
can be used to at least qualitatively describe lateral 
self-diffusion in lipid bilayers~\cite[]{Alm92}. According 
to free area theory, lateral diffusion of a lipid or sterol 
in a bilayer is restricted by the occurrence of a free area greater than 
some critical area adjacent to the diffusing molecule. 
A diffusing molecule spends a comparatively long time---of the 
order of tens of nanoseconds~\cite[]{Tie97,Vat03}---in a cage 
formed by its neighbors, and then, given a large enough 
activation energy and an adjacent free area, jumps.

More specifically, free area theory predicts that the 
lateral diffusion coefficient of a lipid or sterol diffusing 
in a bilayer depends on the free area and the packing properties
as follows~\cite[]{Alm92}:
\begin{equation} 
\label{eq_free_area_theory} 
D_T \sim \exp(-a_0 / a_\mathrm{f}).
\end{equation} 
Here $a_0$ is an estimate for the average cross-sectional area 
for a DPPC or cholesterol molecule and $a_\mathrm{f}$ is a measure 
for the average amount of free area per molecule in the bilayer.

To examine the validity of Eq.~(\ref{eq_free_area_theory}) we 
compute the lateral diffusion coefficients for
DPPC and cholesterol molecules at different cholesterol concentrations.
The lateral tracer diffusion coefficients can be computed 
using the following Einstein relation 
\begin{equation} 
\label{eq_einstein_lateral} 
D_\mathrm{T} = \lim_{t \to \infty} \frac{1}{4 t N_\mathrm{species}} 
      \sum_{i=1}^{N_\mathrm{species}} 
      \langle [ \vec{r}_i(t) - \vec{r}_i(0) ]^2 \rangle. 
\end{equation}
Here $\vec{r}_i(t)$ is the CM position of molecule $i$ at time $t$
and the sum is over all molecules of a given species. 
The lateral diffusion coefficients have been calculated by 
following the position of each molecule in the upper 
(lower) monolayer with respect to the 
center of mass position of the corresponding upper (lower) 
monolayer. Thus, should there be any drift, the motion of the CM of 
each monolayer has been taken into account.

\begin{figure}[h]
\centering
\includegraphics[width=\columnwidth,bb=105 295 425 535,scale=1,clip=true]{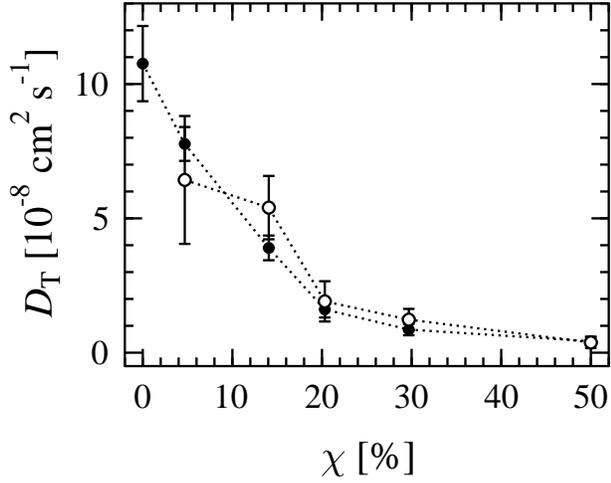}
\caption{Lateral diffusion coefficients of DPPC ($\bullet$) and cholesterol ($\circ$)
molecules as functions of cholesterol concentration.
}
\label{fig_lateral_diffusion}
\end{figure}

Results for lateral diffusion coefficients 
are shown in Fig.~\ref{fig_lateral_diffusion}. The lateral diffusion
coefficients for both DPPC and cholesterol decrease
monotonically with an increasing cholesterol content. This reduction
is qualitatively consistent with experiments~\cite[]{Alm92,Fil03,Kon92}. 
Quantitative comparisons should preferably be made to experimental
techniques that probe lateral diffusion of individual molecules
at time scales comparable to those reached in MD simulations. 
Fluorescence correlation spectroscopy (FCS) measurements should hence
give us a good reference. In FCS measurements for 
DLPC\,/\,cholesterol systems Korlach et al. found that when 
the cholesterol concentration was increased from 0\,\% to 60\,\%.,
$D_\mathrm{T}$ for DLPC was reduced by a factor of ten \cite[]{Kor99}.
Even though the acyl chains of DLPC molecules are shorter than 
those of DPPC molecules, our findings are in reasonable 
accord with Korlach's experiments.

Let us now consider the implications of our results to free 
area theory for lateral diffusion. In free area theory, 
the critical area $a_0$ is essentially a number describing the
close-packed cross-sectional molecular area of the diffusant.
In the same spirit, the average free area per molecule 
$a_\mathrm{f}$ should be characterized by a single number.
We have, however, seen that the free areas per molecule and 
the areas per DPPC and cholesterol molecules are functions
of the distance from the bilayer center. Hence,
it seems that a two-dimensional mean-field model
might be a too simplistic means of describing lateral diffusion.

In our opinion, one should at least not expect free area theory 
to yield quantitative results. It might, however, give qualitative 
predictions about trends in cases where e.\,g.~the cholesterol 
content in a bilayer is increased. With this in mind, 
let us assume that the cholesterol concentration
in a DPPC\,/\,cholesterol bilayer rises from 4.7\,\% to 29.7\,\%.
If we now use the largest possible values for the fractions 
$a_0 / a_\mathrm{f}$, free area theory will give us
upper bounds for the reduction of the lateral diffusion coefficients. 
The lateral diffusion coefficient
for DPPC should according to free area theory  
be now reduced by a factor three at most,
and $D_\mathrm{T}$ for cholesterol should decrease by 
a factor of two. As a matter of fact, the lateral diffusion 
coefficients for both DPPC and cholesterol computed from the 
simulation data are reduced much more strongly, see 
Fig.~\ref{fig_lateral_diffusion}. We can conclude 
that Eq.~(\ref{eq_free_area_theory}) tends to underestimate 
the changes in the values of the lateral diffusion coefficients.

Even though the discrepancies in the predictions of 
Eq.~(\ref{eq_free_area_theory}) and the computed lateral diffusion 
coefficients do exist, we cannot immediately declare free area 
theory incomplete. There is a detail that has been overlooked in 
our discussion so far, and the significance of this detail will 
now be considered. To jump to an adjacent empty site, 
a diffusing molecule needs energy to overcome an activation 
barrier. In free area theory this is accounted for by letting the 
lateral diffusion coefficient be proportional to a Boltzmann factor 
$\exp(-E_\mathrm{a} / k_\mathrm{B} T)$, where $E_\mathrm{a}$ is 
the activation barrier. As a growing cholesterol concentration 
increases the ordering of the DPPC tails and therefore reduces 
the area per molecule, it seems reasonable to expect that 
$E_\mathrm{a}$ should increase with the cholesterol content. 
Experimental results~\cite[]{Alm92} do support this idea but 
are partly contradicting. This is, however, probably due to the 
fitting procedure used \cite[]{Alm92}. In a more recent study, 
Filippov et al. used NMR to study the lateral diffusion in 
palmitoyloleoylphosphocholine\,/\,cholesterol and 
dioleoylphosphocholine\,/\,cholesterol bilayers over a 
cholesterol concentration range of $\sim$0\,--\,45\,mol\,\% 
\cite{Fil03b}. At small $\chi$, they found the apparent (Arrhenius) 
diffusion barrier to be approximately constant, while for 
large $\chi$ the diffusion barrier increased markedly. 
Hence, the neglect of the energy term might in our case lead 
to slight underestimates for the reduction of the lateral 
diffusion coefficients.

Summarizing, we have found that free area theory
correctly predicts the reduction of the lateral diffusion 
coefficients with an increasing cholesterol concentration.
At the same time it seems unnecessary to aim for a quantitative
description with such a simple framework. Instead of being 
based on mean-field arguments, a full theoretical description 
of lateral diffusion should account for local free volume 
fluctuations in the vicinity of diffusing molecules. Atomic-scale 
MD studies in this direction should be feasible in the near future.

\subsection{Area Compressibility} 

Lateral diffusion is clearly influenced by the average amount 
of free area in the bilayer. However, not only the average free 
area, but also fluctuations in the amount of free area should 
play a role here. Recall that free area theory states that 
a diffusion jump is not possible unless there is a large enough 
free area next to the diffusant~\cite[]{Alm92}. Large enough free 
areas are a result of fluctuations, and hence we would expect 
diffusion to depend on the magnitude of the fluctuations: decreasing
fluctuations and slowed lateral diffusion should be coupled.
For similar reasons, it is likely that permeation of 
molecules across membranes can at least partially
be explained by area fluctuations in membranes.

We may quantify area fluctuations in different regions of 
the membrane as follows. The starting point is the average 
occupied area $\langle A_\mathrm{occ}(z) \rangle$, 
i.\,e.,~the area which is not free but occupied by DPPC, cholesterol, 
or water molecules. The occupied area obviously varies with the
distance from the bilayer center $z$. Based on the
occupied area we define an area compressibility as follows:
\begin{equation} \label{eq_acompr} 
\kappa_\mathrm{A}(z) \equiv k_\mathrm{B} T 
     \frac{\langle A_\mathrm{occ}(z) \rangle}
    {\langle \delta A_\mathrm{occ}^2(z) \rangle}. 
\end{equation}
Here $k_\mathrm{B}$ is the Boltzmann constant and 
$\langle \delta A_\mathrm{occ}^2 \rangle = 
\langle A_\mathrm{occ}^2 \rangle - 
\langle A_\mathrm{occ} \rangle^2$. The area compressibility 
is a measure of the fluctuations in the occupied area: a high 
compressibility indicates small fluctuations and a low 
compressibility, correspondingly, large fluctuations. Hence, 
the area compressibility should be related to the 
permeation of small solutes, as well as for the lateral 
diffusion of lipids and sterols.

\begin{figure}[h]
\centering
\includegraphics[width=\columnwidth,bb=105 295 425 535,scale=1,clip=true]{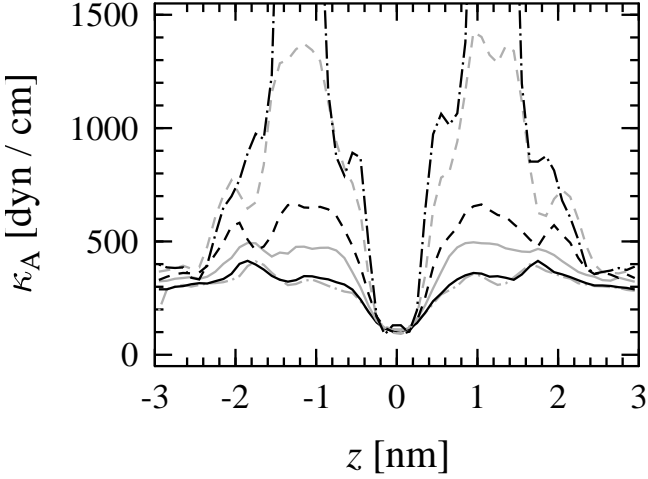}
\caption{Area compressibilities as functions of distance from 
bilayer center. The curves correspond to the cholesterol 
concentrations as indicated in Fig.~\ref{fig_eden_tot}. 
The errors are between 10 and 20\,\%.
}
\label{fig_acompr}
\end{figure}

Figure~\ref{fig_acompr} shows the area compressibility 
profiles computed for systems with different amounts of 
cholesterol. Before focusing on the behavior of 
$\kappa_\mathrm{A}(z)$, let us stress that these 
quantities are sensitive to force 
fields and various computational details and can only be 
used for discussing qualitative trends with an increasing 
cholesterol content.

Regardless of the cholesterol content, all compressibility 
profiles have a minimum in the bilayer center. Moreover, 
the values of compressibility are identical in the center. 
The situation in the water phase makes sense: we expect 
that the compressibilities, irrespective of the cholesterol 
content, should be approximately similar. The interesting 
regions are the tail and head ones. The compressibility 
profiles show two maxima between the bilayer center and 
the water phase, the first at approximately 1\,nm from the 
bilayer center and the second at $1.7 - 2.0$\,nm, depending 
on the cholesterol concentration. Between these we observe 
a local minimum. For pure DPPC and at low cholesterol 
concentrations there is a very flat, plateau-like maximum 
centered at 1\,nm. With more cholesterol, the maximum 
grows considerably. Returning to Fig.~\ref{fig_eden}(f), 
we note that the position of the growing maximum coincides 
with the location of the cholesterol ring structure. 
Therefore we can conclude that the cholesterol steroid rings 
strongly reduce the area fluctuations in the bilayer. From 
the point of view of free area theory, the region with the 
ordered DPPC tails and cholesterol rings seems to be the 
rate-limiting region for lateral diffusion of lipids and sterols.

The local minimum between the two maxima moves from a distance 
1.5\,nm from the bilayer center to 1.9\,nm from the center. 
This means that the minimum is located in a part of the 
bilayer with the uppermost tail methylene groups and parts 
of headgroups. The densities of cholesterol backbone are 
quite small here, while the electron densities of the 
cholesterol hydroxyl groups peak (data not shown). Very few 
cholesterol rings, and hence tails with less order than at 
1\,nm, and possibly also the interface between hydrophobic 
and hydrophilic parts, lead to slightly larger area fluctuations 
here. This could have consequences for the permeation of small 
molecules. Knowing that there are larger area fluctuations 
in this region than elsewhere, does not, however, tell us 
how permeation is affected. Jedlovszky et al., e.\,g., 
found in a recent simulation study of DMPC\,/\,cholesterol 
that the region with the cholesterol hydroxyl groups is indeed 
important from the point of view of permeation~\cite[]{Jed03b}. 
The effect on the actual rate of the permeation process, 
nevertheless, must depend on the properties of the 
permeant molecule.

\subsection{Close-Packed Areas from Ordering of Acyl Chains?
\label{sect_ordering_area}}

Let us return to the average area per DPPC and investigate 
whether anything can be said about the close-packed area of
a DPPC molecule based on the chain order parameters. Traditionally, the 
use of deuterium NMR experiments to determine the average 
area per DPPC has resulted in a wide variety of 
values~\cite[]{Nag00}. This is not so much due to the underlying 
results for the order parameters as due to the interpretation 
of the results. Petratche et al.~have rather recently suggested 
a way of relating the deuterium order parameter to the average 
chain travel distance along the bilayer normal~\cite[]{Pet99}:
\begin{equation} \label{eq_order_parameter_and_travel}
\langle D_n \rangle = \frac{D_\mathrm{M}}{2} \left( 
1 + \sqrt{\frac{-8 S_\mathrm{CD}^n - 1}
{3}}\right).
\end{equation}
Here $\langle D_n \rangle$ is the average chain travel distance 
along the bilayer normal for segment $n$, $D_\mathrm{M}$ the maximum 
travel per methylene for all-trans chains oriented perpendicularly 
to the bilayer, and $S_\mathrm{CD}^n$ the deuterium order parameter for 
segment $n$. By assuming that $\langle A_n \rangle \approx 
V_\mathrm{CH_2} / \langle D_n \rangle$, where $V_\mathrm{CH_2}$ 
is the volume per methylene group 
(this is only true if $\langle D_n^2\rangle \approx 
\langle D_n \rangle^2$, see~\cite[]{Pet99}) and recalling 
Eq.~(\ref{eq_deuterium_order_parameter}), we may write: 
\begin{equation} \label{eq_order_parameter_and_area}
\frac{1}{2} S_\mathrm{zz}^n = \frac{1}{8} + \frac{3}{8} 
\left( \frac{2 A_0}{A_n} + 1 \right)^2,
\end{equation}
where $A_0$ is the area occupied by a fully ordered 
phospholipid molecule. By examination of Figs.~\ref{fig_s_vs_n} 
and~\ref{fig_area_per_num}(a), we can conclude that it is 
unrealistic to expect that Eq.~(\ref{eq_order_parameter_and_area}) 
should allow one to extract the detailed form of 
$a_\mathrm{DPPC}(z)$ from $S_\mathrm{zz}$. Nevertheless, 
Eq.~(\ref{eq_order_parameter_and_area}) might be useful in 
predicting the average areas per DPPC molecule in the tail region, 
e.\,g.~at a distance 1\,nm from the bilayer center, where the 
headgroup density is negligible for all cholesterol 
concentrations.

\begin{figure}[h]
\centering
\includegraphics[width=\columnwidth,bb=105 295 425 535,scale=1,clip=true]{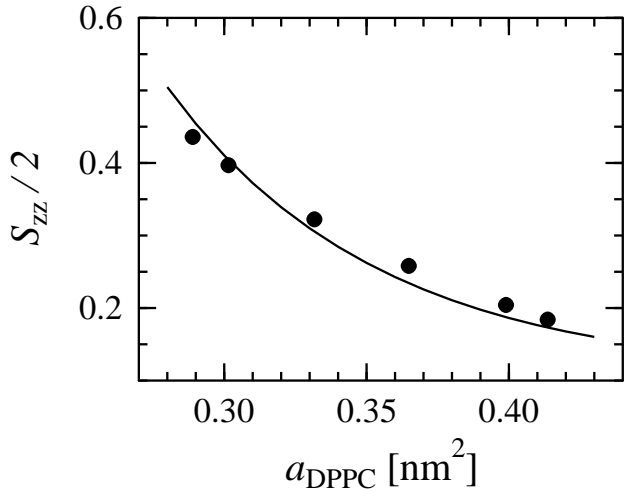}
\caption{Order parameters vs.~close-packed areas for DPPC at 1\,nm from 
bilayer center. The markers represent values computed from 
the simulations and the solid 
line is a fit to these data based on 
Eq.~(\ref{eq_order_parameter_and_area}).
}
\label{fig_s_vs_a}
\end{figure}

To find the values of the order parameters at 1\,nm from the 
bilayer center, we use electron density profiles calculated 
separately for each methylene group in the hydrocarbon tails 
(data not shown) to determine which segment is located at 
a distance 1\,nm from the center for each cholesterol 
concentration separately. The order parameters at 1\,nm are 
then calculated as averages over the segments whose electron 
density profiles peak at the close vicinity of 1\,nm and over 
the {\it sn}--1 and {\it sn}--2 tails. The close-packed areas 
for DPPC at 1\,nm from the center, in turn, can be easily 
obtained from the $a_\mathrm{DPPC}(z)$ profiles. The resulting 
values and a fit to Eq.~(\ref{eq_order_parameter_and_area})  
are shown in Fig.~\ref{fig_s_vs_a}. The fit is astonishingly 
good, given that Eq.~(\ref{eq_order_parameter_and_area}) has 
been developed for a pure phospholipid bilayer and is based 
on a rather simple model. The best fit is obtained with 
$A_0 \approx 0.28$\,nm$^2$.  $A_0$ should in this case be 
interpreted as the area occupied by the fully ordered 
{\it sn}--1 and {\it sn}--2 tails. Hence, the agreement 
with Fig.~\ref{fig_area_per_num}(a) is surprisingly good. 
Yet one should not pay too much attention to the exact numerical 
value here, as it probably depends on the details of the force 
field. 

Hofs\"a{\ss} et al.~have used Eq.~(\ref{eq_order_parameter_and_area}) 
in a slightly different setting~\cite[]{Hof03}. As order parameters 
they have used averages over the order parameter profiles from 
segment 3 to segment 8, and the average areas per DPPC they used 
contain some free area. They, too, find that 
Eq.~(\ref{eq_order_parameter_and_area}) gives a very good fit to their 
data. However, we expect that order parameters are related 
to close-packed cross-sectional areas for DPPC chains rather 
than to average areas per DPPC containing an arbitrary 
amount of free area.

\section{Summary and Conclusions}

We have performed 100\,ns molecular dynamics simulations at 
$T = 323$\,K on a pure DPPC bilayer and composite 
DPPC\,/\,cholesterol bilayers with 4.7\,\%, 12.5\,\%, 
20.3\,\%, 29.7\,\%, and 50.0\,\% cholesterol. The main focus 
has been on the packing of molecules, free area in different
parts of the bilayer, and lateral diffusion of DPPC and cholesterol
molecules. Especially the interplay between these properties has been
considered.

To investigate the packing and free area properties, we have
introduced a novel method for estimating the average space 
occupied by DPPC, cholesterol, and water molecules, along with 
the average amount of free space, in different regions of the bilayer. 
Using this method we have computed the average cross-sectional 
areas for DPPC and cholesterol, as well as the total free area, 
as functions of the distance from the bilayer center. The 
method should be generally applicable for all kinds of pure
and composite bilayers. Moreover, it could be used for 
investigating bilayers with integral proteins and in such 
a way finding out how the bilayer structure is changed in the 
vicinity of embedded proteins.

Inspection of the cross-sectional close-packed area profiles 
for DPPC and cholesterol, i.\,e.,~the close-packed areas as 
functions of the distance from the bilayer center, has shown 
that cholesterol alters the packing of molecules and reduces 
the amount of occupied space. These phenomena have been 
quite generally explained in terms of the form of the cholesterol 
molecule and the ordering effect of cholesterol on parts of 
the phospholipid tails.

Cholesterol has also been found to significantly reduce the 
average amount of free space in all regions of the bilayer. 
We have further discovered that the form of the free area 
profiles, i.\,e.,~the average amount of free area as a function 
of the distance from the bilayer center, is altered. These 
changes seem to reflect the ones observed in the close-packed 
area profiles for DPPC and cholesterol. We therefore conclude 
that the packing and free area properties are strongy coupled.

Also lateral diffusion of DPPC and cholesterol molecules has 
been found to be strongly reduced with an increasing cholesterol 
content. Further, the changes in the packing properties and the 
average amount of free area seem to be reflected in the behavior 
of the lateral diffusion coefficients for DPPC and cholesterol 
molecules. We have, however, learned that even though so-called 
free area theories correctly predict the suppressed lateral 
diffusion with reduced free area, the dependence cannot be 
quantitatively described by mean-field models such as free area 
theory. Not only are the average free areas or volumes relevant 
for diffusion, but also the size distribution, shape, and local 
fluctuations of the free volumes in the bilayer are important. 
It would hence be interesting to see how cholesterol influences 
the size distribution of free volumes in the bilayer.

\bigskip

\section{Acknowledgments}
This work has, in part, been supported by the Academy of Finland 
through its Center of Excellence Program (E.\,F. and I.\,V.), 
the National Graduate School in Materials Physics (E.\,F.), 
the Academy of Finland Grant Nos.~54113, 00119 (M.\,K.), 80246 (I.\,V.),
and 80851 (M.\,H.), and the Jenny and Antti Wihuri Foundation (M.\,H.).
M.\,P. would like to acknowledge the support through
the Marie Curie fellowship No.~HPMF--CT--2002--01794. 
We would also like to thank the Finnish IT Center for Science 
and the HorseShoe (DCSC) supercluster computing facility 
at the University of Southern Denmark for computer resources. 
Finally, we are grateful to Ole G. Mouritsen and Peter Lindqvist 
for fruitful discussions.

\end{document}